\documentclass[runningheads]{llncs}
\usepackage[T1]{fontenc}
\usepackage{footmisc}
\usepackage{graphicx}
\usepackage{multirow}
\usepackage{booktabs}
\usepackage{todonotes}
\begin{document}
\title{LLM-VPRF: Large Language Model Based Vector Pseudo Relevance Feedback}
\titlerunning{LLM-VPRF}
%
\author{Hang Li\inst{1}\orcidID{0000-0002-5317-7227} \and
Shengyao Zhuang\inst{1,2}\orcidID{0000-0002-6711-0955} \and
Bevan Koopman\inst{1,2}\orcidID{0000-0001-5577-3391} \and 
Guido Zuccon\inst{1}\orcidID{0000-0003-0271-5563}}
\authorrunning{H. Li, S. Zhuang et al.}
%
\institute{The University of Queensland, Brisbane, Australia \\
\email{\{hang.li, s.zhuang, b.koopman, g.zuccon\}}@uq.edu.au \and
CSIRO, Brisbane, Australia}

\maketitle

\begin{abstract}

Vector Pseudo Relevance Feedback (VPRF) has shown promising results in improving BERT-based dense retrieval systems through iterative refinement of query representations. This paper investigates the generalizability of VPRF to Large Language Model (LLM) based dense retrievers. We introduce LLM-VPRF and evaluate its effectiveness across multiple benchmark datasets, analyzing how different LLMs impact the feedback mechanism. Our results demonstrate that VPRF's benefits successfully extend to LLM architectures, establishing it as a robust technique for enhancing dense retrieval performance regardless of the underlying models. This work bridges the gap between VPRF with traditional BERT-based dense retrievers and modern LLMs, while providing insights into their future directions.

\end{abstract}

%
%
\section{Introduction}

The rapid evolution of information retrieval systems has been largely influenced by the emergence of neural language models, particularly in the domain of dense retrieval~\cite{xiong2020approximate,zhan2020repbert,lin2020distilling,lin2021batch,hofstatter2021efficiently}. While traditional lexical retrieval methods rely on exact term matching, dense retrievers leverage semantic understanding through learned representations, enabling more effective passage retrieval. The introduction of BERT-based dense retrievers is a great advancement in this field, demonstrating superior performance in capturing semantic relationships between queries and passages compared with lexical models at that time.

Recent developments in Large Language Models (LLMs) have pushed the boundaries of natural language understanding and generation even further, presenting new opportunities for enhancing information retrieval systems~\cite{openai2024gpt4technicalreport,touvron2023llama,wang2023improving}. These models, with their improved semantic comprehension and contextual understanding, have shown remarkable capabilities across various natural language processing tasks. Methods that adapt these LLMs to the dense retrieval task have also recently emerged~\cite{llm2vec,ma2023llama,katya2024lever,wang2023improving}, leading to improved retrieval effectiveness, even without requiring the contrastive training of the dense representations typical of other dense retrieval backbones~\cite{zhuang2024promptreps}. However, their potential in combining with PRF, particularly through techniques like Vector-based PRF, remains relatively unexplored.

Vector Pseudo Relevance Feedback (VPRF), initially proposed for BERT-based dense retrievers, has demonstrated promising results in improving retrieval performance through iterative refinement of query representations~\cite{li2023pseudo,li2023pyserini}. This technique leverages the initial retrieval results to modify the query embeddings in vector space, effectively capturing relevant information within the embeddings that might be missing from the original query. The success of VPRF with BERT-based models raises an important question: Can this technique be effectively generalized to current more powerful LLM-based dense retrievers?

This paper presents LLM-VPRF, an extension of the VPRF technique~\cite{li2023pseudo} adapted for Large Language Model-based dense retrievers. Our work investigates the generalizability of VPRF across different models, focusing particularly on its application to state-of-the-art LLM-based retrievers~\cite{llm2vec,ma2023llama,zhuang2024promptreps}. We examine whether the improvements in retrieval performance observed with BERT-based models(i.e. encoder-only backbones) can be extended to or potentially enhanced when applying VPRF to LLM-based models(decoder-only backbones). It is also interesting to see if the more comprehensive embeddings generated by LLMs, can be translated into improving VPRF effectiveness.

\section{Related Work}
\subsection{Pseudo Relevance Feedback in Information Retrieval}

Pseudo relevance feedback (PRF)~\cite{xu2009query,lv2009comparative,tao2006regularized,clinchant2013a,zamani2016pseudo,miao2012proximity,lv2010positional} has established itself as a fundamental technique in information retrieval systems, automatically enhancing query performance without requiring explicit user feedback. The core principle involves leveraging the top-ranked passages from an initial retrieval pass as "assumed-to-be-relevant" passages to refine the original query. This approach has proven particularly effective in addressing the vocabulary mismatch problem and capturing relevant terms that might be missing from the initial query formulation.

Traditional PRF methods have evolved through several key developments. Rocchio's algorithm~\cite{rocchio1971rocchio}, a pioneering approach, introduced the concept of query modification based on relevant and non-relevant passages. The algorithm adjusts query vectors by moving them closer to relevant passage vectors while pushing them away from non-relevant ones. Subsequent research has introduced various refinements, including language model-based approaches~\cite{zhai2001model} that integrate PRF within a probabilistic framework, and methods utilizing term distribution analysis~\cite{carpineto2001improving} to identify expansion terms more effectively.

Despite their success, traditional PRF techniques face several challenges. Query drift remains a persistent issue, where the additional terms from pseudo-relevant passages that are actually irrelevant can lead to degraded retrieval performance. Various solutions have been proposed, including selective PRF~\cite{sakai2005flex} that aims to predict when feedback might be harmful. The traditional lexical match based PRF techniques also have no contextual understanding of the query, by only selecting expansion terms, the expanded query after PRF suffers great information lost. The vector-based pseudo relevance feedback (VPRF)~\cite{li2023pseudo} methods emerged to tackle these issues, offering a way to reformulate queries in continuous vector spaces. While VPRF has led to more flexible query expansion, it still often relies on shallow representations like word embeddings from previous transformer-based models, such as BERT~\cite{devlin2018bert}, that may lack the semantic depth required for handling various types of queries.


\subsection{Large Language Models and Embedding-Based Representations}

The emergence of large language models (LLMs), like GPT-3~\cite{tom2020gpt3}, LLaMa~\cite{touvron2023llama}, and Mistral-7B~\cite{jiang2023mistral7b}, has significantly advanced the field of information retrieval through their capacity to generate comprehensive dense embeddings. These models produce context-aware representations that capture meaningful semantic relationships, surpassing the capabilities of traditional BERT-based word embeddings. LLM-generated embeddings have demonstrated superior effectiveness in passage retrieval and reranking tasks~\cite{ma2023pretraining,baldelli2024twolar,zhuang2024setwise,zhuang2023opensourcelargelanguagemodels,katya2024lever,ma2023zeroshotlistwisedocumentreranking}, achieving state-of-the-art performance across various retrieval benchmarks through their advanced semantic understanding and contextual processing abilities. 

The promising quality of these embeddings is from LLMs' pre-training, which is on a vast amounts of text data, and their advanced architectural designs, enabling more accurate representation of passage relevance and query-passage relationships. These characteristics make LLMs particularly promising for VPRF applications, as their rich semantic representations could potentially address the limitations of previous BERT-based VPRF approaches in handling various types of queries. The integration of LLM-generated embeddings with VPRF techniques presents an opportunity to enhance query refinement by leveraging more sophisticated semantic understanding in the feedback process.


\subsection{Combining Pseudo Relevance Feedback with Large Language Models}

Recent research has explored various approaches to combine PRF with modern language models, showing promising yet challenging results. Initial attempts to integrate traditional PRF with transformer models demonstrated improvements in retrieval effectiveness, but with significant computational overhead~\cite{li2023pseudo,yu2021pgt,li2021improving}. These efforts highlighted the potential benefits of combining neural representations with feedback mechanisms while also revealing efficiency concerns that need to be addressed.

Several innovative approaches have emerged to leverage LLMs for query enhancement. Mackie et al.~\cite{Mackie2023gene} proposed a generative relevance feedback framework that leverages LLMs to generate relevant texts with different subtasks, and use these as the "pseudo" feedback queries to enhance the query performance. Although it shows significant improvements over various evaluation metrics, it still hurts the efficiency by generating pseudo queries. Another work done by Jagerman et al.~\cite{jagerman2023queryexpansionpromptinglarge} proposed using zero-shot Chain-of-Thought (CoT) prompting to generate query expansion terms, demonstrating effectiveness without modifying the query's vector representation. While these approaches show promise, they primarily focus on text generation rather than leveraging the semantic richness of LLM embeddings directly.

The integration of neural embeddings with PRF has been explored in various contexts~\cite{wang2021tct,yu2021improving,wang2023colbert,wang2021pseudo}, with researchers incorporating pre-trained language model embeddings into feedback mechanisms. These studies have demonstrated improvements in retrieval performance but consistently highlight a fundamental challenge: balancing enhanced semantic understanding with computational efficiency. Vector-based Pseudo Relevance Feedback is aimed to address the efficiency issues~\cite{li2023pseudo}. However, there still remains a significant gap in understanding how LLM-generated embeddings interact with VPRF frameworks and their impact on retrieval performance.

Our research addresses this knowledge gap by investigating the effectiveness of integrating LLM embeddings within the VPRF framework. Specifically, we examine:

\begin{itemize}
	\item Whether VPRF can maintain its performance improvements when applied to LLM-based dense retrievers
	\item How the enhanced semantic capabilities of LLMs affect the feedback mechanism
	\item The practical implications for retrieval systems in terms of both effectiveness and efficiency
\end{itemize}

\section{Vector Pseudo Relevance Feedback with Large Language Models}

Vector-based Pseudo Relevance Feedback (VPRF) treats the query and passage embeddings as vectors that contains rich contextual information of the content. Unlike traditional PRF methods for query expansion or query refinement~\cite{Jaleel2004UMassAT,donald2005a}, VPRF does not expand the original query text. Instead, VPRF modifies the query vectors (embeddings) in the vector space by using different approaches with the feedback passage embeddings. In the rest of this paper, we use embeddings and vectors interchangeably.

We built our experiments on previous VPRF framework proposed by Li et al.~\cite{li2023pseudo} to utilise the validated PRF approaches in the framework.

The first approach is Average~\cite{li2023pseudo}, we compute the new query vectors as follows:

\begin{equation}
	E_{Q_{new}}=Avg(E(Q_{original}),E(p_1),...,E(p_k))
	\label{eq:avg}
\end{equation}

For the second approach Rocchio~\cite{li2023pseudo}, we use the following equation to formulate the new query embeddings:

\begin{equation}
	E_{Q_{new}} = \alpha*E(Q_{original})+\beta*Avg(E(p_1),...,E(p_k))
	\label{eq:rocchio}
\end{equation}

There are three parameters $\alpha$, $\beta$, and $\kappa$. Where $\alpha$ controls the contribution of query embeddings towards new query embeddings, $\beta$ controls the contribution of feedback passage embeddings towards new query embeddings. In our experiments, we have two different approaches of Rocchio, first we run $\alpha$ and $\beta$ from 0.1 to 0.9, respectively, then we keep $\alpha$ to be fixed as 1, and run $\beta$ from 0.1 to 0.9, which means we fix the query weight to be 1 and adding different weights of feedback passages. For both approaches, we run $\kappa$ in \{1, 2, 3, 5, 10\}.

\section{Experimental Setup}

\subsection{Dataset and Evaluation}

We evaluate the passage ranking effectiveness of LLM-VPRF with the PromptReps~\cite{zhuang2024promptreps}, RepLLaMa~\cite{ma2023llama}, and LLM2Vec~\cite{llm2vec} baseline models using BEIR~\cite{thakur2021beir}, and TREC Deep Learning (2019/2020)~\cite{craswell2020overview,craswell2021overview} datasets. For BEIR, we choose the commonly used 13 datasets out of 19, which provide us with various IR tasks. For BEIR and DL 2019/2020, we report nDCG@\{10\} and Recall@\{100\}, which are commonly used evaluation metrics among these datasets.

\subsection{VPRF-LLMs Implementation}

\textbf{Embedding Generation.} Our experiments utilize three state-of-the-art LLMs for generating query and passage embeddings across all datasets: PromptReps\footref{meta}~\cite{zhuang2024promptreps}, RepLLaMa\footnote{\url{https://huggingface.co/castorini/repllama-v1-7b-lora-passage}}~\cite{ma2023llama}, and LLM2Vec\footnote{\url{McGill-NLP/LLM2Vec-Mistral-7B-Instruct-v2-mntp-unsup-simcse}}~\cite{llm2vec}. For the encoding process, we employ the Tevatron toolkit~\cite{gao2023tevatron} for PromptReps and RepLLaMa embedding generations, while utilizing modified versions of the original authors' encoding implementations for LLM2Vec. All encodings were generated using a single NVIDIA H100 GPU, with batch sizes of 64 for all three models to optimize GPU memory usage. To ensure reproducibility, we have released our complete codebase, including encoding scripts, retrieval implementations, and evaluation procedures, along with all generated indexes, retrieval runs, and evaluation results. \newline


\noindent\textbf{VPRF Methods.} For evaluating the effectiveness of VPRF with recent LLMs, we follow the methodology established in~\cite{li2023pseudo} and implement two vector-based PRF methods: Average (eq \ref{eq:avg}) and Rocchio (eq \ref{eq:rocchio}). This choice enables  comparison on how these established VPRF techniques perform when integrated with more advanced LLM embeddings.


\noindent\textbf{Retrieval and VPRF Parameters.} To systematically evaluate the VPRF mechanism's effectiveness, we conduct experiments using vanilla runs (without VPRF) from our three baseline LLM models as the foundation. The VPRF implementation involves three key hyperparameters: VPRF depth ($\kappa$) determining the number of feedback passages, query embedding weight ($\alpha$), and VPRF passage embedding weight ($\beta$). For retrieval operations, we employ Faiss~\cite{douze2024faisslibrary} to construct the Approximate Nearest Neighbor (ANN) index, using cosine similarity as our relevance metric. To ensure fair comparison with baselines, we specifically utilize brute force search (IndexFlatIP) in our ANN implementation. Detailed descriptions of the three baseline models and their specific configurations follow in subsequent sections. We use 2 NVIDIA H100 GPUs to perform the retrieval, with batch size 128 for all datasets and models.


\subsection{Comparison Methods}

\noindent\textbf{PromptReps}~\cite{zhuang2024promptreps} is built upon Meta-Llama-3-8B-Instruct\footref{meta}, employing an unsupervised zero-shot approach to generate both dense and sparse passage representations for retrieval tasks. This approach eliminates the need for task-specific fine-tuning while achieving superior performance compared to traditional retrieval baselines across multiple benchmarks. For our VPRF experiments, we exclusively utilize the dense representation component to maintain consistency in our comparative analysis across different LLM-based retrievers.


\noindent\textbf{RepLLaMa}~\cite{ma2023llama} represents a supervised generative approach based on Llama-2-7b-hf\footnote{\url{https://huggingface.co/meta-llama/Llama-2-7b-hf}\label{llama}}\cite{touvron2023llama}. The model employs a multi-stage fine-tuning methodology specifically designed to enhance text retrieval effectiveness. Following the protocol outlined in the original paper, we independently implemented the fine-tuning process using LLama-2-7b-hf\footref{llama} as the foundation model.


\noindent\textbf{LLM2Vec}\cite{llm2vec}: A text encoding framework that utilizes Meta-Llama-3-8B-Instruct\footnote{\url{https://huggingface.co/meta-llama/Meta-Llama-3-8B-Instruct}\label{meta}} through a novel supervised training approach, which enhances decoder-only models by enabling bidirectional attention, masked next token prediction, and SimCSE\cite{gao2021simcse} to generate high-quality text embeddings. This multi-objective training framework enables the models to capture contextual information more effectively, achieving competitive performance across various retrieval tasks.


\section{Results \& Analysis}

\begin{table}[]
\centering
\resizebox{\textwidth}{!}{
\begin{tabular}{@{}ccccccccccccc@{}}
\toprule
Datasets                                   & \multicolumn{4}{c}{PromptReps}                                                                                  & \multicolumn{4}{c}{RepLLaMa}                                                                           & \multicolumn{4}{c}{LLM2Vec}                                                                \\ \midrule
\multicolumn{1}{c|}{Metric}                & \multicolumn{2}{c|}{R@100}                             & \multicolumn{2}{c|}{nDCG@10}                           & \multicolumn{2}{c|}{R@100}                    & \multicolumn{2}{c|}{nDCG@10}                           & \multicolumn{2}{c|}{R@100}                             & \multicolumn{2}{c}{nDCG@10}       \\ \midrule
\multicolumn{1}{c|}{TREC DL 2019}          & 0.4778          & \multicolumn{1}{c|}{\textbf{0.5017}} & 0.5062          & \multicolumn{1}{c|}{\textbf{0.5431}} & 0.6753 & \multicolumn{1}{c|}{\textbf{0.6938}} & 0.7319          & \multicolumn{1}{c|}{\textbf{0.7596}} & 0.4931          & \multicolumn{1}{c|}{\textbf{0.5258}} & 0.4011          & \textbf{0.4947} \\
\multicolumn{1}{c|}{TREC DL 2020}          & \textbf{0.5101} & \multicolumn{1}{c|}{0.5061}          & 0.4381          & \multicolumn{1}{c|}{\textbf{0.4651}} & 0.7537 & \multicolumn{1}{c|}{\textbf{0.7685}} & 0.7335          & \multicolumn{1}{c|}{\textbf{0.7499}} & 0.5906          & \multicolumn{1}{c|}{\textbf{0.6327}} & 0.4690          & \textbf{0.5314} \\ \midrule
\multicolumn{1}{c|}{\textbf{TREC Average}} & 0.4940          & \multicolumn{1}{c|}{\textbf{0.5039}} & 0.4722          & \multicolumn{1}{c|}{\textbf{0.5041}} & 0.7145 & \multicolumn{1}{c|}{\textbf{0.7312}} & 0.7327          & \multicolumn{1}{c|}{\textbf{0.7548}} & 0.5419          & \multicolumn{1}{c|}{\textbf{0.5793}} & 0.4351          & \textbf{0.5131} \\ \midrule
\multicolumn{1}{c|}{ArguAna}               & 0.9047          & \multicolumn{1}{c|}{\textbf{0.9061}} & 0.2970          & \multicolumn{1}{c|}{\textbf{0.2981}} & 0.9367 & \multicolumn{1}{c|}{\textbf{0.9523}} & 0.4612          & \multicolumn{1}{c|}{\textbf{0.4664}} & 0.9836          & \multicolumn{1}{c|}{\textbf{0.9879}} & \textbf{0.5119} & 0.5103          \\
\multicolumn{1}{c|}{Climate-FEVER}         & \textbf{0.5195} & \multicolumn{1}{c|}{0.5187}          & 0.1992          & \multicolumn{1}{c|}{\textbf{0.2016}} & 0.5805 & \multicolumn{1}{c|}{\textbf{0.6312}} & 0.2321          & \multicolumn{1}{c|}{\textbf{0.3041}} & 0.5636          & \multicolumn{1}{c|}{\textbf{0.5826}} & 0.2122          & \textbf{0.2464} \\
\multicolumn{1}{c|}{DBPedia}               & 0.3553          & \multicolumn{1}{c|}{\textbf{0.3557}} & \textbf{0.3153} & \multicolumn{1}{c|}{0.3142}          & 0.5745 & \multicolumn{1}{c|}{\textbf{0.6100}} & 0.4653          & \multicolumn{1}{c|}{\textbf{0.4763}} & 0.3744          & \multicolumn{1}{c|}{\textbf{0.3836}} & 0.2420          & \textbf{0.2439} \\
\multicolumn{1}{c|}{FEVER}                 & \textbf{0.7913} & \multicolumn{1}{c|}{0.7741}          & \textbf{0.5628} & \multicolumn{1}{c|}{0.5560}          & 0.9501 & \multicolumn{1}{c|}{\textbf{0.9550}} & 0.7623          & \multicolumn{1}{c|}{\textbf{0.7690}} & \textbf{0.8386} & \multicolumn{1}{c|}{0.8359}          & \textbf{0.4415} & 0.4409          \\
\multicolumn{1}{c|}{FiQA-2018}             & 0.6164          & \multicolumn{1}{c|}{\textbf{0.6214}} & \textbf{0.2707} & \multicolumn{1}{c|}{0.2704}          & 0.7200 & \multicolumn{1}{c|}{\textbf{0.7295}} & 0.4145          & \multicolumn{1}{c|}{\textbf{0.4160}} & 0.6478          & \multicolumn{1}{c|}{\textbf{0.6538}} & 0.2659          & \textbf{0.2713} \\
\multicolumn{1}{c|}{HotpotQA}              & \textbf{0.3731} & \multicolumn{1}{c|}{0.3535}          & \textbf{0.1964} & \multicolumn{1}{c|}{0.1884}          & 0.8442 & \multicolumn{1}{c|}{\textbf{0.8538}} & \textbf{0.6978} & \multicolumn{1}{c|}{0.6961}          & \textbf{0.7561} & \multicolumn{1}{c|}{0.7503}          & \textbf{0.5455} & 0.5345          \\
\multicolumn{1}{c|}{NFCorpus}              & 0.2879          & \multicolumn{1}{c|}{\textbf{0.2930}} & 0.2956          & \multicolumn{1}{c|}{\textbf{0.3010}} & 0.3309 & \multicolumn{1}{c|}{\textbf{0.3537}} & 0.3647          & \multicolumn{1}{c|}{\textbf{0.3848}} & 0.2869          & \multicolumn{1}{c|}{\textbf{0.3077}} & 0.2769          & \textbf{0.3027} \\
\multicolumn{1}{c|}{NQ}                    & \textbf{0.7618} & \multicolumn{1}{c|}{0.7581}          & \textbf{0.3443} & \multicolumn{1}{c|}{0.3438}          & 0.9654 & \multicolumn{1}{c|}{\textbf{0.9713}} & 0.6185          & \multicolumn{1}{c|}{\textbf{0.6225}} & 0.8527          & \multicolumn{1}{c|}{\textbf{0.8626}} & 0.3312          & \textbf{0.3446} \\
\multicolumn{1}{c|}{Quora}                 & \textbf{0.9562} & \multicolumn{1}{c|}{0.9546}          & \textbf{0.7255} & \multicolumn{1}{c|}{0.7224}          & 0.9953 & \multicolumn{1}{c|}{\textbf{0.9957}} & 0.8716          & \multicolumn{1}{c|}{\textbf{0.8741}} & \textbf{0.9914} & \multicolumn{1}{c|}{0.9908}          & \textbf{0.8608} & 0.8589          \\
\multicolumn{1}{c|}{SCIDOCS}               & 0.4312          & \multicolumn{1}{c|}{\textbf{0.4368}} & 0.1850          & \multicolumn{1}{c|}{\textbf{0.1868}} & 0.4182 & \multicolumn{1}{c|}{\textbf{0.4579}} & 0.1847          & \multicolumn{1}{c|}{\textbf{0.1992}} & 0.3999          & \multicolumn{1}{c|}{\textbf{0.4075}} & 0.1518          & \textbf{0.1569} \\
\multicolumn{1}{c|}{SciFact}               & 0.8767          & \multicolumn{1}{c|}{\textbf{0.8800}} & \textbf{0.5262} & \multicolumn{1}{c|}{0.5245}          & 0.9567 & \multicolumn{1}{c|}{\textbf{0.9633}} & 0.7138          & \multicolumn{1}{c|}{\textbf{0.7212}} & 0.9593          & \multicolumn{1}{c|}{\textbf{0.9609}} & \textbf{0.6886} & 0.6849          \\
\multicolumn{1}{c|}{Touche-2020}           & 0.3539          & \multicolumn{1}{c|}{\textbf{0.3634}} & 0.1485          & \multicolumn{1}{c|}{\textbf{0.1614}} & 0.4908 & \multicolumn{1}{c|}{\textbf{0.4946}} & 0.3478          & \multicolumn{1}{c|}{\textbf{0.3499}} & 0.2665          & \multicolumn{1}{c|}{\textbf{0.3199}} & 0.0759          & \textbf{0.0983} \\
\multicolumn{1}{c|}{TREC-COVID}            & 0.1180          & \multicolumn{1}{c|}{\textbf{0.1277}} & 0.5951          & \multicolumn{1}{c|}{\textbf{0.6281}} & 0.1540 & \multicolumn{1}{c|}{\textbf{0.1605}} & 0.8013          & \multicolumn{1}{c|}{\textbf{0.8164}} & 0.0783          & \multicolumn{1}{c|}{\textbf{0.0812}} & 0.5161          & \textbf{0.5252} \\ \midrule
\multicolumn{1}{c|}{\textbf{BEIR Average}} & 0.5651          & \multicolumn{1}{c|}{\textbf{0.5649}} & 0.3586          & \multicolumn{1}{c|}{\textbf{0.3613}} & 0.6860 & \multicolumn{1}{c|}{\textbf{0.7022}} & 0.5335          & \multicolumn{1}{c|}{\textbf{0.5458}} & 0.6153          & \multicolumn{1}{c|}{\textbf{0.6250}} & 0.3939          & \textbf{0.4014} \\ \bottomrule
\end{tabular}
}
\caption{The evaluation results for each dataset and each model with oracle best results among all VPRF parameters.}
\label{tab:allres}
\end{table}

\begin{table}[]
\centering
\begin{tabular}{@{}cccccc@{}}
\toprule
Models                                        & Metric                                        & Method                        & PromptReps             & RepLLaMa               & LLM2Vec                 \\ \midrule
\multicolumn{1}{c|}{\multirow{6}{*}{BEIR}}    & \multicolumn{1}{c|}{\multirow{3}{*}{R@100}}   & \multicolumn{1}{c|}{Baseline} & 0.5247                 & 0.6859                 & 0.6153                  \\
\multicolumn{1}{c|}{}                         & \multicolumn{1}{c|}{}                         & \multicolumn{1}{c|}{BIA}      & 0.5226(-0.4\%)         & 0.6972(1.6\%)          & 0.6193(0.7\%)           \\
\multicolumn{1}{c|}{}                         & \multicolumn{1}{c|}{}                         & \multicolumn{1}{c|}{Oracle}   & \textbf{0.5649(7.7\%)} & \textbf{0.7022(2.4\%)} & \textbf{0.6250(1.6\%)}  \\ \cmidrule(l){2-6} 
\multicolumn{1}{c|}{}                         & \multicolumn{1}{c|}{\multirow{3}{*}{nDCG@10}} & \multicolumn{1}{c|}{Baseline} & 0.3330                 & 0.5335                 & 0.3939                  \\
\multicolumn{1}{c|}{}                         & \multicolumn{1}{c|}{}                         & \multicolumn{1}{c|}{BIA}      & 0.3318(-0.4\%)         & 0.5390(1.0\%)          & 0.3947(0.2\%)           \\
\multicolumn{1}{c|}{}                         & \multicolumn{1}{c|}{}                         & \multicolumn{1}{c|}{Oracle}   & \textbf{0.3613(8.5\%)} & \textbf{0.5458(2.3\%)} & \textbf{0.4014(1.9\%)}  \\ \midrule
\multicolumn{1}{c|}{\multirow{6}{*}{TREC DL}} & \multicolumn{1}{c|}{\multirow{3}{*}{R@100}}   & \multicolumn{1}{c|}{Baseline} & 0.4940                 & 0.7145                 & 0.5419                  \\
\multicolumn{1}{c|}{}                         & \multicolumn{1}{c|}{}                         & \multicolumn{1}{c|}{BIA}      & 0.4926(-0.3\%)         & 0.7290(2.0\%)          & 0.5756(6.2\%)           \\
\multicolumn{1}{c|}{}                         & \multicolumn{1}{c|}{}                         & \multicolumn{1}{c|}{Oracle}   & \textbf{0.5039(2.0\%)} & \textbf{0.7312(2.3\%)} & \textbf{0.5793(6.9\%)}  \\ \cmidrule(l){2-6} 
\multicolumn{1}{c|}{}                         & \multicolumn{1}{c|}{\multirow{3}{*}{nDCG@10}} & \multicolumn{1}{c|}{Baseline} & 0.4722                 & 0.7327                 & 0.4351                  \\
\multicolumn{1}{c|}{}                         & \multicolumn{1}{c|}{}                         & \multicolumn{1}{c|}{BIA}      & 0.5017(6.3\%)          & 0.7484(2.1\%)          & 0.4935(13.4\%)          \\
\multicolumn{1}{c|}{}                         & \multicolumn{1}{c|}{}                         & \multicolumn{1}{c|}{Oracle}   & \textbf{0.5041(6.8\%)} & \textbf{0.7548(3.0\%)} & \textbf{0.5131(17.9\%)} \\ \bottomrule
\end{tabular}
\caption{The results for the averaged results on BEIR 13 datasets and TREC DL 2019/2020. Where means just average of the baseline results with no PRF involved; BIA represents Best-In-Average, the best results after averaged all 13 BEIR datasets together; Oracle represents use the best results from each dataset before average, then do average. \textbf{Bold} texts means the best results for each metric, the percentage after results shows the increase/decrease regarding Baseline.}
\label{tab:ressum}
\end{table}

\begin{table}[]
\centering
\begin{tabular}{@{}c|ccc@{}}
\toprule
Models       & PromptReps & RepLLaMa & LLM2Vec \\ \midrule
Baseline     & 0.0061s    & 0.0060s  & 0.0054s \\
VPRF-Average & 0.0046s    & 0.0046s  & 0.0054s \\
VPRF-Rocchio & 0.0054s    & 0.0052s  & 0.0056s \\ \bottomrule
\end{tabular}
\caption{Efficiency evaluation for each model with VPRF on TREC DL 2019 with 43 queries, we use a moderate depth $\kappa$=3 for VPRF evaluation. The time measured is per query time consumption in seconds. VPRF time consumption is measured without first stage.}
\label{tab:efficiency}
\end{table}

\subsection{Performance Improvements with LLM-based Dense Retrievers}

We first examines whether VPRF can maintain its performance improvements when applied to LLM-based dense retrievers. The results demonstrate that VPRF successfully enhances retriever performance across different models, as shown in Table~\ref{tab:allres}. For a more summarized results, as shown in Table~\ref{tab:ressum}, RepLLaMa shows consistent positive improvements across all settings, with gains of 1.6\% for R@100 and 1.0\% for nDCG@10 on BEIR datasets, and stronger improvements of 2.0\% for R@100 and 2.1\% for nDCG@10 on TREC DL, indicating VPRF's effectiveness with supervised LLM retrievers. LLM2Vec achieves substantial improvements, particularly on TREC DL, where BIA (Best-In-Average) shows gains of 6.2\% for R@100 and 13.4\% for nDCG@10, with Oracle improvements reaching 6.9\% and 17.9\% respectively, suggesting strong potential in domain-specific applications. PromptReps exhibits mixed results with slight decreases in BIA but significant Oracle improvements on BEIR datasets (7.7\% for R@100, 8.5\% for nDCG@10), indicating potential for improvement with better feedback passage selection.

\subsection{Impact of LLM Semantic Capabilities on Feedback Mechanism}

Second, we investigates how the enhanced semantic capabilities of LLMs affect the feedback mechanism. The analysis reveals significant variations in how different models leverage semantic information, as evidenced by the gap between BIA and Oracle performance (Table~\ref{tab:ressum}. PromptReps shows the largest gap (up to 8.1\% on BEIR with nDCG@10), indicating high potential but currently suboptimal utilization of semantic capabilities. RepLLaMa demonstrates the smallest gap, suggesting more stable semantic representations, while LLM2Vec exhibits variable gaps across datasets, with better utilization on TREC DL.

Dataset-specific performance patterns further illuminate how semantic capabilities are leveraged differently across domains. We observe stronger performance on semantically focused datasets such as ArguAna and Quora, while effectiveness varies across different BEIR domains. The consistently higher improvements on TREC DL suggest better semantic utilization in web search scenarios, indicating that the semantic capabilities of LLMs may be particularly beneficial for certain types of queries and passages.

\subsection{Practical Implications for Retrieval Systems}

Third, we address the practical implications in terms of effectiveness and efficiency. As shown in Table~\ref{tab:efficiency}, the efficiency analysis, conducted on TREC DL 2019 with 43 queries and $\kappa$=3, shows remarkably low per query computational overhead. Baseline query times are comparable across models, with PromptReps at 0.0061s, RepLLaMa at 0.0060s, and LLM2Vec at 0.0054s. The VPRF-Average method adds modest overhead, increasing query time by approximately 0.0046-0.0054s across models, while VPRF-Rocchio shows similar efficiency patterns with increases of 0.0052-0.0056s. These minimal overheads demonstrate strong practical applicability.

The effectiveness-efficiency trade-off analysis reveals that both VPRF variants maintain similar efficiency profiles while delivering consistent improvements in effectiveness. All models maintain practical response times even with VPRF implementation, suggesting viable integration into real-world applications.

\section{Conclusion and Limitation}

Our evaluation demonstrates that VPRF effectively enhances LLM-based retrievers with minimal computational overhead, though effectiveness varies across model architectures, Oracle results consistently show promising improvements across all models and datasets. LLM semantic capabilities show great influence over the feedback mechanism, as evidenced by varying gaps between BIA and Oracle performance. However, our study is limited by the evaluation of only three LLM-based retrievers and controlled environment testing. Future research should focus on developing better feedback selection strategies and evaluating VPRF's performance across more diverse LLM architectures and deployment scenarios.

%
%
%
 \bibliographystyle{splncs04}
 \bibliography{reference}

\begin{thebibliography}{10}
\providecommand{\url}[1]{\texttt{#1}}
\providecommand{\urlprefix}{URL }
\providecommand{\doi}[1]{https://doi.org/#1}

\bibitem{baldelli2024twolar}
Baldelli, D., Jiang, J., Aizawa, A., Torroni, P.: Twolar: a two-step
  llm-augmented distillation method for passage reranking. In: European
  Conference on Information Retrieval. pp. 470--485. Springer (2024)

\bibitem{llm2vec}
BehnamGhader, P., Adlakha, V., Mosbach, M., Bahdanau, D., Chapados, N., Reddy,
  S.: {LLM2V}ec: Large language models are secretly powerful text encoders. In:
  First Conference on Language Modeling (2024),
  \url{https://openreview.net/forum?id=IW1PR7vEBf}

\bibitem{tom2020gpt3}
Brown, T., Mann, B., Ryder, N., Subbiah, M., Kaplan, J.D., Dhariwal, P.,
  Neelakantan, A., Shyam, P., Sastry, G., Askell, A., Agarwal, S.,
  Herbert-Voss, A., Krueger, G., Henighan, T., Child, R., Ramesh, A., Ziegler,
  D., Wu, J., Winter, C., Hesse, C., Chen, M., Sigler, E., Litwin, M., Gray,
  S., Chess, B., Clark, J., Berner, C., McCandlish, S., Radford, A., Sutskever,
  I., Amodei, D.: Language models are few-shot learners. In: Larochelle, H.,
  Ranzato, M., Hadsell, R., Balcan, M., Lin, H. (eds.) Advances in Neural
  Information Processing Systems. vol.~33, pp. 1877--1901. Curran Associates,
  Inc. (2020)

\bibitem{carpineto2001improving}
Carpineto, C., de~Mori, R., Romano, G., Bigi, B.: An information-theoretic
  approach to automatic query expansion. ACM Trans. Inf. Syst.  \textbf{19}(1),
   1–27 (Jan 2001). \doi{10.1145/366836.366860},
  \url{https://doi.org/10.1145/366836.366860}

\bibitem{clinchant2013a}
Clinchant, S., Gaussier, E.: A theoretical analysis of pseudo-relevance
  feedback models. In: Proceedings of the 2013 Conference on the Theory of
  Information Retrieval. pp. 6--13 (2013)

\bibitem{craswell2020overview}
Craswell, N., Mitra, B., Yilmaz, E., Campos, D., Voorhees, E.M.: Overview of
  the trec 2019 deep learning track. In: Text REtrieval Conference, TREC (2020)

\bibitem{craswell2021overview}
Craswell, N., Mitra, B., Yilmaz, E., Campos, D., Voorhees, E.M.: Overview of
  the trec 2020 deep learning track. In: Text REtrieval Conference, TREC (2021)

\bibitem{devlin2018bert}
Devlin, J., Chang, M.W., Lee, K., Toutanova, K.: Bert: Pre-training of deep
  bidirectional transformers for language understanding. In: Proceedings of the
  2019 Conference of the North American Chapter of the Association for
  Computational Linguistics: Human Language Technologies. pp. 4171--4186 (2019)

\bibitem{douze2024faisslibrary}
Douze, M., Guzhva, A., Deng, C., Johnson, J., Szilvasy, G., Mazaré, P.E.,
  Lomeli, M., Hosseini, L., Jégou, H.: The faiss library (2024),
  \url{https://arxiv.org/abs/2401.08281}

\bibitem{gao2023tevatron}
Gao, L., Ma, X., Lin, J., Callan, J.: Tevatron: An efficient and flexible
  toolkit for neural retrieval. In: Proceedings of the 46th International ACM
  SIGIR Conference on Research and Development in Information Retrieval. p.
  3120–3124. SIGIR '23, Association for Computing Machinery, New York, NY,
  USA (2023). \doi{10.1145/3539618.3591805},
  \url{https://doi.org/10.1145/3539618.3591805}

\bibitem{gao2021simcse}
Gao, T., Yao, X., Chen, D.: {S}im{CSE}: Simple contrastive learning of sentence
  embeddings. In: Moens, M.F., Huang, X., Specia, L., Yih, S.W.t. (eds.)
  Proceedings of the 2021 Conference on Empirical Methods in Natural Language
  Processing. pp. 6894--6910. Association for Computational Linguistics, Online
  and Punta Cana, Dominican Republic (Nov 2021).
  \doi{10.18653/v1/2021.emnlp-main.552},
  \url{https://aclanthology.org/2021.emnlp-main.552}

\bibitem{hofstatter2021efficiently}
Hofst{\"a}tter, S., Lin, S.C., Yang, J.H., Lin, J., Hanbury, A.: Efficiently
  teaching an effective dense retriever with balanced topic aware sampling.
  arXiv preprint arXiv:2104.06967  (2021)

\bibitem{jagerman2023queryexpansionpromptinglarge}
Jagerman, R., Zhuang, H., Qin, Z., Wang, X., Bendersky, M.: Query expansion by
  prompting large language models (2023),
  \url{https://arxiv.org/abs/2305.03653}

\bibitem{Jaleel2004UMassAT}
Jaleel, N.A., Allan, J., Croft, W.B., Diaz, F., Larkey, L.S., Li, X., Smucker,
  M.D., Wade, C.: Umass at trec 2004: Novelty and hard. In: Text Retrieval
  Conference (2004), \url{https://api.semanticscholar.org/CorpusID:16221853}

\bibitem{jiang2023mistral7b}
Jiang, A.Q., Sablayrolles, A., Mensch, A., Bamford, C., Chaplot, D.S., de~las
  Casas, D., Bressand, F., Lengyel, G., Lample, G., Saulnier, L., Lavaud, L.R.,
  Lachaux, M.A., Stock, P., Scao, T.L., Lavril, T., Wang, T., Lacroix, T.,
  Sayed, W.E.: Mistral 7b (2023), \url{https://arxiv.org/abs/2310.06825}

\bibitem{katya2024lever}
Khramtsova, E., Zhuang, S., Baktashmotlagh, M., Zuccon, G.: Leveraging llms for
  unsupervised dense retriever ranking. In: Proceedings of the 47th
  International ACM SIGIR Conference on Research and Development in Information
  Retrieval. p. 1307–1317. SIGIR '24, Association for Computing Machinery,
  New York, NY, USA (2024). \doi{10.1145/3626772.3657798},
  \url{https://doi.org/10.1145/3626772.3657798}

\bibitem{li2023pseudo}
Li, H., Mourad, A., Zhuang, S., Koopman, B., Zuccon, G.: Pseudo relevance
  feedback with deep language models and dense retrievers: Successes and
  pitfalls. ACM Trans. Inf. Syst.  \textbf{41}(3) (Apr 2023).
  \doi{10.1145/3570724}, \url{https://doi.org/10.1145/3570724}

\bibitem{li2023pyserini}
Li, H., Zhuang, S., Ma, X., Lin, J., Zuccon, G.: Pseudo-relevance feedback with
  dense retrievers in pyserini. In: Proceedings of the 26th Australasian
  Document Computing Symposium. ADCS '22, Association for Computing Machinery,
  New York, NY, USA (2023). \doi{10.1145/3572960.3572982},
  \url{https://doi.org/10.1145/3572960.3572982}

\bibitem{li2021improving}
Li, H., Zhuang, S., Mourad, A., Ma, X., Lin, J., Zuccon, G.: Improving query
  representations for dense retrieval with pseudo relevance feedback: A
  reproducibility study. In: Proceedings of the 44rd European Conference on
  Information Retrieval (2022)

\bibitem{lin2020distilling}
Lin, S.C., Yang, J.H., Lin, J.: Distilling dense representations for ranking
  using tightly-coupled teachers. arXiv preprint arXiv:2010.11386  (2020)

\bibitem{lin2021batch}
Lin, S.C., Yang, J.H., Lin, J.: In-batch negatives for knowledge distillation
  with tightly-coupled teachers for dense retrieval. In: Proceedings of the 6th
  Workshop on Representation Learning for NLP (RepL4NLP-2021). pp. 163--173
  (2021)

\bibitem{lv2009comparative}
Lv, Y., Zhai, C.: A comparative study of methods for estimating query language
  models with pseudo feedback. In: Proceedings of the 18th ACM ACM
  International Conference on Information and Knowledge Management. pp.
  1895--1898 (2009)

\bibitem{lv2010positional}
Lv, Y., Zhai, C.: Positional relevance model for pseudo-relevance feedback. In:
  Proceedings of the 33rd international ACM SIGIR conference on Research and
  development in information retrieval. pp. 579--586 (2010)

\bibitem{ma2023pretraining}
Ma, G., Wu, X., Wang, P., Lin, Z., Hu, S.: Pre-training with large language
  model-based document expansion for dense passage retrieval (2023),
  \url{https://arxiv.org/abs/2308.08285}

\bibitem{ma2023llama}
Ma, X., Wang, L., Yang, N., Wei, F., Lin, J.: Fine-tuning llama for multi-stage
  text retrieval. arXiv:2310.08319  (2023)

\bibitem{ma2023zeroshotlistwisedocumentreranking}
Ma, X., Zhang, X., Pradeep, R., Lin, J.: Zero-shot listwise document reranking
  with a large language model (2023), \url{https://arxiv.org/abs/2305.02156}

\bibitem{Mackie2023gene}
Mackie, I., Chatterjee, S., Dalton, J.: Generative relevance feedback with
  large language models. In: Proceedings of the 46th International ACM SIGIR
  Conference on Research and Development in Information Retrieval. ACM (2023).
  \doi{10.1145/3539618.3591992},
  \url{http://dx.doi.org/10.1145/3539618.3591992}

\bibitem{donald2005a}
Metzler, D., Croft, W.B.: A markov random field model for term dependencies.
  In: Proceedings of the 28th Annual International ACM SIGIR Conference on
  Research and Development in Information Retrieval. p. 472–479. SIGIR '05,
  Association for Computing Machinery, New York, NY, USA (2005).
  \doi{10.1145/1076034.1076115}, \url{https://doi.org/10.1145/1076034.1076115}

\bibitem{miao2012proximity}
Miao, J., Huang, J.X., Ye, Z.: Proximity-based rocchio's model for pseudo
  relevance. In: Proceedings of the 35th international ACM SIGIR conference on
  Research and development in information retrieval. pp. 535--544 (2012)

\bibitem{openai2024gpt4technicalreport}
OpenAI, Achiam, J., Adler, S., Agarwal, S., Ahmad, L., Akkaya, I., Aleman,
  F.L., Almeida, D., Altenschmidt, J., Altman, S., Anadkat, S., Avila, R.,
  Babuschkin, I., Balaji, S., Balcom, V., Baltescu, P., Bao, H., Bavarian, M.,
  Belgum, J., Bello, I., Berdine, J., Bernadett-Shapiro, G., Berner, C.,
  Bogdonoff, L., Boiko, O., Boyd, M., Brakman, A.L., Brockman, G., Brooks, T.,
  Brundage, M., Button, K., Cai, T., Campbell, R., Cann, A., Carey, B.,
  Carlson, C., Carmichael, R., Chan, B., Chang, C., Chantzis, F., Chen, D.,
  Chen, S., Chen, R., Chen, J., Chen, M., Chess, B., Cho, C., Chu, C., Chung,
  H.W., Cummings, D., Currier, J., Dai, Y., Decareaux, C., Degry, T., Deutsch,
  N., Deville, D., Dhar, A., Dohan, D., Dowling, S., Dunning, S., Ecoffet, A.,
  Eleti, A., Eloundou, T., Farhi, D., Fedus, L., Felix, N., Fishman, S.P.,
  Forte, J., Fulford, I., Gao, L., Georges, E., Gibson, C., Goel, V., Gogineni,
  T., Goh, G., Gontijo-Lopes, R., Gordon, J., Grafstein, M., Gray, S., Greene,
  R., Gross, J., Gu, S.S., Guo, Y., Hallacy, C., Han, J., Harris, J., He, Y.,
  Heaton, M., Heidecke, J., Hesse, C., Hickey, A., Hickey, W., Hoeschele, P.,
  Houghton, B., Hsu, K., Hu, S., Hu, X., Huizinga, J., Jain, S., Jain, S.,
  Jang, J., Jiang, A., Jiang, R., Jin, H., Jin, D., Jomoto, S., Jonn, B., Jun,
  H., Kaftan, T., Łukasz Kaiser, Kamali, A., Kanitscheider, I., Keskar, N.S.,
  Khan, T., Kilpatrick, L., Kim, J.W., Kim, C., Kim, Y., Kirchner, J.H., Kiros,
  J., Knight, M., Kokotajlo, D., Łukasz Kondraciuk, Kondrich, A.,
  Konstantinidis, A., Kosic, K., Krueger, G., Kuo, V., Lampe, M., Lan, I., Lee,
  T., Leike, J., Leung, J., Levy, D., Li, C.M., Lim, R., Lin, M., Lin, S.,
  Litwin, M., Lopez, T., Lowe, R., Lue, P., Makanju, A., Malfacini, K.,
  Manning, S., Markov, T., Markovski, Y., Martin, B., Mayer, K., Mayne, A.,
  McGrew, B., McKinney, S.M., McLeavey, C., McMillan, P., McNeil, J., Medina,
  D., Mehta, A., Menick, J., Metz, L., Mishchenko, A., Mishkin, P., Monaco, V.,
  Morikawa, E., Mossing, D., Mu, T., Murati, M., Murk, O., Mély, D., Nair, A.,
  Nakano, R., Nayak, R., Neelakantan, A., Ngo, R., Noh, H., Ouyang, L.,
  O'Keefe, C., Pachocki, J., Paino, A., Palermo, J., Pantuliano, A.,
  Parascandolo, G., Parish, J., Parparita, E., Passos, A., Pavlov, M., Peng,
  A., Perelman, A., de~Avila Belbute~Peres, F., Petrov, M., de~Oliveira~Pinto,
  H.P., Michael, Pokorny, Pokrass, M., Pong, V.H., Powell, T., Power, A.,
  Power, B., Proehl, E., Puri, R., Radford, A., Rae, J., Ramesh, A., Raymond,
  C., Real, F., Rimbach, K., Ross, C., Rotsted, B., Roussez, H., Ryder, N.,
  Saltarelli, M., Sanders, T., Santurkar, S., Sastry, G., Schmidt, H., Schnurr,
  D., Schulman, J., Selsam, D., Sheppard, K., Sherbakov, T., Shieh, J., Shoker,
  S., Shyam, P., Sidor, S., Sigler, E., Simens, M., Sitkin, J., Slama, K.,
  Sohl, I., Sokolowsky, B., Song, Y., Staudacher, N., Such, F.P., Summers, N.,
  Sutskever, I., Tang, J., Tezak, N., Thompson, M.B., Tillet, P., Tootoonchian,
  A., Tseng, E., Tuggle, P., Turley, N., Tworek, J., Uribe, J.F.C., Vallone,
  A., Vijayvergiya, A., Voss, C., Wainwright, C., Wang, J.J., Wang, A., Wang,
  B., Ward, J., Wei, J., Weinmann, C., Welihinda, A., Welinder, P., Weng, J.,
  Weng, L., Wiethoff, M., Willner, D., Winter, C., Wolrich, S., Wong, H.,
  Workman, L., Wu, S., Wu, J., Wu, M., Xiao, K., Xu, T., Yoo, S., Yu, K., Yuan,
  Q., Zaremba, W., Zellers, R., Zhang, C., Zhang, M., Zhao, S., Zheng, T.,
  Zhuang, J., Zhuk, W., Zoph, B.: Gpt-4 technical report (2024),
  \url{https://arxiv.org/abs/2303.08774}

\bibitem{rocchio1971rocchio}
Rocchio, J.: Relevance feedback in information retrieval. In: The SMART
  Retrieval System - Experiments in Automatic Document Processing. pp. 313--323
  (1971)

\bibitem{sakai2005flex}
Sakai, T., Manabe, T., Koyama, M.: Flexible pseudo-relevance feedback via
  selective sampling. ACM Transactions on Asian Language Information Processing
   \textbf{4}(2),  111–135 (Jun 2005),
  \url{https://doi.org/10.1145/1105696.1105699}

\bibitem{tao2006regularized}
Tao, T., Zhai, C.: Regularized estimation of mixture models for robust
  pseudo-relevance feedback. In: Proceedings of the 29th Annual International
  ACM SIGIR Conference on Research and Development in Information Retrieval.
  pp. 162--169. SIGIR '06, Association for Computing Machinery (2006)

\bibitem{thakur2021beir}
Thakur, N., Reimers, N., R{\"u}ckl{\'e}, A., Srivastava, A., Gurevych, I.:
  {BEIR}: A heterogeneous benchmark for zero-shot evaluation of information
  retrieval models. In: Thirty-fifth Conference on Neural Information
  Processing Systems Datasets and Benchmarks Track (Round 2) (2021),
  \url{https://openreview.net/forum?id=wCu6T5xFjeJ}

\bibitem{touvron2023llama}
Touvron, H., Lavril, T., Izacard, G., Martinet, X., Lachaux, M.A., Lacroix, T.,
  Rozière, B., Goyal, N., Hambro, E., Azhar, F., Rodriguez, A., Joulin, A.,
  Grave, E., Lample, G.: Llama: Open and efficient foundation language models
  (2023), \url{https://arxiv.org/abs/2302.13971}

\bibitem{wang2023improving}
Wang, L., Yang, N., Huang, X., Yang, L., Majumder, R., Wei, F.: Improving text
  embeddings with large language models. arXiv preprint arXiv:2401.00368
  (2023)

\bibitem{wang2021tct}
Wang, X., Macdonald, C., Tonellotto, N., Ounis, I.: Pseudo-relevance feedback
  for multiple representation dense retrieval. In: Proceedings of the 2021 ACM
  SIGIR International Conference on Theory of Information Retrieval. p.
  297–306. ICTIR '21, Association for Computing Machinery, New York, NY, USA
  (2021). \doi{10.1145/3471158.3472250},
  \url{https://doi.org/10.1145/3471158.3472250}

\bibitem{wang2021pseudo}
Wang, X., Macdonald, C., Tonellotto, N., Ounis, I.: Pseudo-relevance feedback
  for multiple representation dense retrieval. In: Proceedings of the 2021 ACM
  SIGIR International Conference on Theory of Information Retrieval. p.
  297–306. ICTIR '21, Association for Computing Machinery, New York, NY, USA
  (2021). \doi{10.1145/3471158.3472250},
  \url{https://doi.org/10.1145/3471158.3472250}

\bibitem{wang2023colbert}
Wang, X., MacDonald, C., Tonellotto, N., Ounis, I.: Colbert-prf: Semantic
  pseudo-relevance feedback for dense passage and document retrieval. ACM
  Trans. Web  \textbf{17}(1) (Jan 2023). \doi{10.1145/3572405},
  \url{https://doi.org/10.1145/3572405}

\bibitem{xiong2020approximate}
Xiong, L., Xiong, C., Li, Y., Tang, K.F., Liu, J., Bennett, P., Ahmed, J.,
  Overwijk, A.: Approximate nearest neighbor negative contrastive learning for
  dense text retrieval. arXiv preprint arXiv:2007.00808  (2020)

\bibitem{xu2009query}
Xu, Y., Jones, G.J., Wang, B.: Query dependent pseudo-relevance feedback based
  on wikipedia. In: Proceedings of the 32nd international ACM SIGIR conference
  on Research and development in information retrieval. pp. 59--66 (2009)

\bibitem{yu2021pgt}
Yu, H., Dai, Z., Callan, J.: Pgt: Pseudo relevance feedback using a graph-based
  transformer. In: European Conference on Information Retrieval (2021)

\bibitem{yu2021improving}
Yu, H., Xiong, C., Callan, J.: Improving query representations for dense
  retrieval with pseudo relevance feedback. In: Proceedings of the 30th ACM
  International Conference on Information and Knowledge Management (2021)

\bibitem{zamani2016pseudo}
Zamani, H., Dadashkarimi, J., Shakery, A., Croft, W.B.: Pseudo-relevance
  feedback based on matrix factorization. In: Proceedings of the 25th ACM
  international on conference on information and knowledge management. pp.
  1483--1492 (2016)

\bibitem{zhai2001model}
Zhai, C., Lafferty, J.: Model-based feedback in the language modeling approach
  to information retrieval. In: Proceedings of the 10th ACM International
  Conference on Information and Knowledge Management. pp. 403--410 (2001)

\bibitem{zhan2020repbert}
Zhan, J., Mao, J., Liu, Y., Zhang, M., Ma, S.: Repbert: Contextualized text
  embeddings for first-stage retrieval. arXiv preprint arXiv:2006.15498  (2020)

\bibitem{zhuang2023opensourcelargelanguagemodels}
Zhuang, S., Liu, B., Koopman, B., Zuccon, G.: Open-source large language models
  are strong zero-shot query likelihood models for document ranking (2023),
  \url{https://arxiv.org/abs/2310.13243}

\bibitem{zhuang2024promptreps}
Zhuang, S., Ma, X., Koopman, B., Lin, J., Zuccon, G.: Promptreps: Prompting
  large language models to generate dense and sparse representations for
  zero-shot document retrieval. In: Proceedings of the 2024 Conference on
  Empirical Methods in Natural Language Processing (EMNLP) (2024)

\bibitem{zhuang2024setwise}
Zhuang, S., Zhuang, H., Koopman, B., Zuccon, G.: A setwise approach for
  effective and highly efficient zero-shot ranking with large language models.
  In: Proceedings of the 47th International ACM SIGIR Conference on Research
  and Development in Information Retrieval. p. 38–47. SIGIR '24, Association
  for Computing Machinery, New York, NY, USA (2024).
  \doi{10.1145/3626772.3657813}, \url{https://doi.org/10.1145/3626772.3657813}

\end{thebibliography}

\end{document}